# Evaluation of Noise Reduction Methods for Sentence Recognition by Sinhala Speaking Listeners

Malitha Gunawardhana, Chathuki Navanjana, Dinithi Fernando, Nipuna Upeksha, Anjula De Silva
*Department of Electronic and Telecommunication Engineering, University of Moratuwa*
Sri Lanka

*Abstract*—Noise reduction is a crucial aspect of hearing aids, which researchers have been striving to address over the years. However, most existing noise reduction algorithms have primarily been evaluated using English. Considering the linguistic differences between English and Sinhala languages, including variation in syllable structures and vowel duration, it is very important to assess the performance of noise reduction tailored to the Sinhala language. This paper presents a comprehensive analysis between wavelet transformation and adaptive filters for noise reduction in Sinhala languages. We investigate the performance of ten wavelet families with soft and hard thresholding methods against adaptive filters with Normalized Least Mean Square, Least Mean Square Average Normalized Least Mean Square, Recursive Least Square, and Adaptive Filtering Averaging optimization algorithms along with cepstral and energy-based voice activity detection algorithms. The performance evaluation is done using objective metrics; Signal to Noise Ratio (SNR) and Perceptual Evaluation of Speech Quality (PESQ) and a subjective metric; Mean Opinion Score (MOS). A newly recorded Sinhala language audio dataset and the NOIZEUS database by the University of Texas, Dallas were used for the evaluation. Our code is available at https://github.com/ChathukiKet/Evaluation-of-Noise-Reduction-Methods

## I. INTRODUCTION

The main objective of hearing aids is to enhance speech signals while minimizing background noise, making noise reduction a crucial factor for improving auditory experiences. When choosing a noise reduction algorithm, it is important to consider the right balance between computational complexity and resource requirements.For instance, the spectral subtraction method offers a lower complexity compared to other algorithms but can result in higher residual noise due to excessive resource utilization [1]. On the other hand, the subspace algorithm with Wiener filters requires less resources but has a high complexity [1].

Wavelet transforms and adaptive filtering are widely used methods for noise reduction. In wavelet transformation, wavelet coefficients are calculated using the convolution between the observed signals and a wavelet, providing localization in both the time domain and frequency domain [2]–[4]. The most critical factor in this method is setting a threshold for noise suppression with minimal distortions to the speech signal [5]–[8]. In adaptive filtering, an optimization algorithm is used to reduce the error between the noise signal and the clean signal and a Voice Activity Detection (VAD) algorithm is used to identify the presence of the voice. Least Mean Square (LMS) [9], [10], Normalized Least Mean Square (NLMS) [10]–[12], Recursive Least Square (RLS) [9]–[11], [13], Adaptive Filtering Averaging (AFA) [12], [13], Average Normalized Least Mean Square (ANLMS) [12] are the most popular optimizing algorithms while cepstral based VAD [14], [15] and energy-based VAD [16] methods being the most popular voice activity detection methods. The performance of adaptive filtering algorithms depends on the complexity, convergence, and stability of the selected optimization algorithm and the performance of the selected VAD algorithm.

Latest Machine Learning models can learn to differentiate speech from noise when trained with features of speech and noise separately. Deep Learning models use different architectures of Artificial Neural Networks to learn from data [17]. Feed-forward Neural Networks [17], Recurrent Neural Networks [17], and Deep Belief Networks are the most popular Deep Neural Network models that are used in speech enhancements. Even though Machine Learning methods work well, there are limitations as those are computationally expensive and require large datasets for training. [18]–[20].

Furthermore, the role of linguistic characteristics in the performance of noise reduction algorithms is often overlooked. In particular, this research focuses on the Sinhala language, primarily spoken in Sri Lanka. The linguistic aspects of Sinhala greatly differ from English, with unique variations in syllable structures, vowel duration, and rhythm. The different rhythmic and syllable patterns may affect the behaviour of noise reduction algorithms. Additionally, the Sinhala language exhibits significantly different phonetic and phonological features compared to English. Notably, the bulk of noise reduction research has been conducted in English and other widely spoken languages. As a consequence, the performance of noise reduction techniques in languages such as Sinhala has not been thoroughly investigated. This paper aims to fill this gap by providing a comprehensive analysis of wavelet transformation and adaptive filters for noise reduction in the Sinhala language. By doing so, the research hopes to advance the field of hearing aid technology and noise

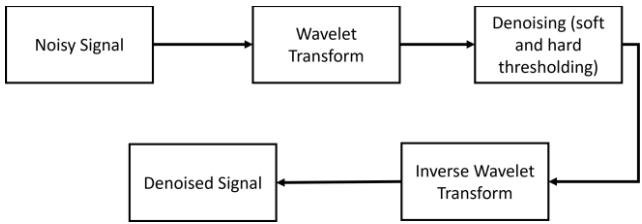

Fig. 1: General approach of denoising a signal by applying thresholds on wavelet coefficients

reduction methods, ensuring more inclusivity by catering to a broader range of languages.

## II. METHODOLOGY

In this paper, initially, we applied wavelet packet transform using multiple wavelet families with different thresholding methods and adaptive filters with different optimization and VAD methods on the NOIZEUS database [21]. MATLAB software was used to implement algorithms. Even though we used objective measures to compare the performance of these algorithms, a contradiction was observed on the perceived clarity of these denoised speech signals when listening. Hence, to obtain a subjective comparison, a new dataset was recorded in Sinhala language (so that it can be tested with local participants) with the guidance of an audiologist.

### A. Wavelet Transform based denoising algorithm

Wavelet transform is used to map time domain signals into the wavelet space which includes frequency information preserving its time information. Therefore, it can be used to compute the wavelet energy of a particular signal for a given time and a scale [2], [5]. Fig. 1 shows the general approach of denoising a signal with wavelet transform.

We considered DWT and WPT approaches for applying the wavelet transform in the discrete domain. We decompose a given signal into five levels with DWT and WPT using the following wavelet families; Haar, Daubechies − db5, db10, db15, Symlets − sym5, sym10, sym15 and Coiflets − coif3, coif4. In each case, the wavelet was convoluted with the input signal. If the similarity was strong between the input signal and the wavelet, a higher coefficient was obtained.

The threshold value for denoising was obtained using balance sparsity norms and fixed-form universal threshold methods. The balance sparsity norms method returned a threshold such that the percentages of retained energy and the number of zeros in the signal are the same [22]. Following the determination of thresholds, the application of thresholds was performed both with soft and hard thresholding methods. Finally, the inverse wavelet transform was applied to reconstruct the denoised audio signal.

### B. Adaptive Filtering

Adaptive filters are digital filters that can self-adjust their coefficients using an optimizing algorithm to give the best match to the desired signal while minimizing the error between the input signal and the desired signal. The main idea is to suppress the noise of the input noisy signal and output a denoised signal, with minimal distortion to the speech quality [9]. Fig. 2 shows the adaptive filtering approach used for noise cancellation. Here, $x(n)$ is the reference signal, $d(n)$ is the desired signal, $e(n)$ is the error signal, $y(n)$ is the output signal, and $\mu$ is the convergence parameter.

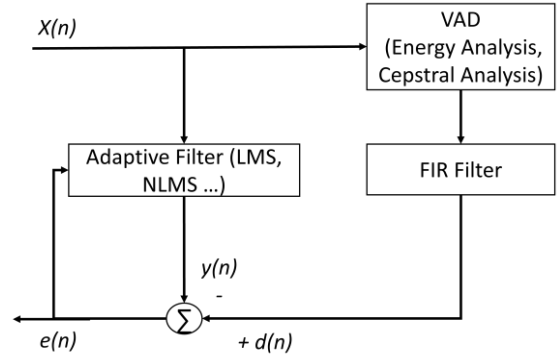

Fig. 2: Noise cancellation using adaptive filtering method

It follows a linear adaptive filtering method, and a VAD approach has been used to estimate the desired signal. We have considered two VAD methods; one based on energy analysis and the other based on cepstral analysis. As the optimizing algorithm, five methods were considered: LMS, NLMS, ANLMS, RLS, and AFA. The best VAD approach and the optimizing algorithm was chosen after analyzing their performance in subjective and objective metrics. All the computations were done as [11].

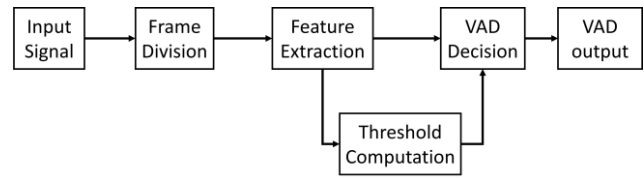

Fig. 3: VAD block diagram

*1) Optimizing algorithms:* When using an adaptive filter, an optimizing algorithm has to be chosen by considering the factors of computational requirements, rate of convergence, robustness, etc. [10]. We implement several optimizing algorithms and among them NLMS outperformed other algorithms. These optimizing algorithms differ from each other according to the way they are used for updating their weights ($w_n$), while the error signal calculation and output signal calculation are the same. Weight updating equations for LMS,

$$w_{n+1} = w_n + \mu e(n)x(n) \quad (1)$$

where $\mu = 0.09$ and order = 46. For NLMS,

$$w_{n+1} = w_n + \frac{\alpha}{c + ||x(n)||^2} e(n)x(n) \quad (2)$$

where c = 0.01, $\alpha = 0.09$ and order = 70. For RLS,

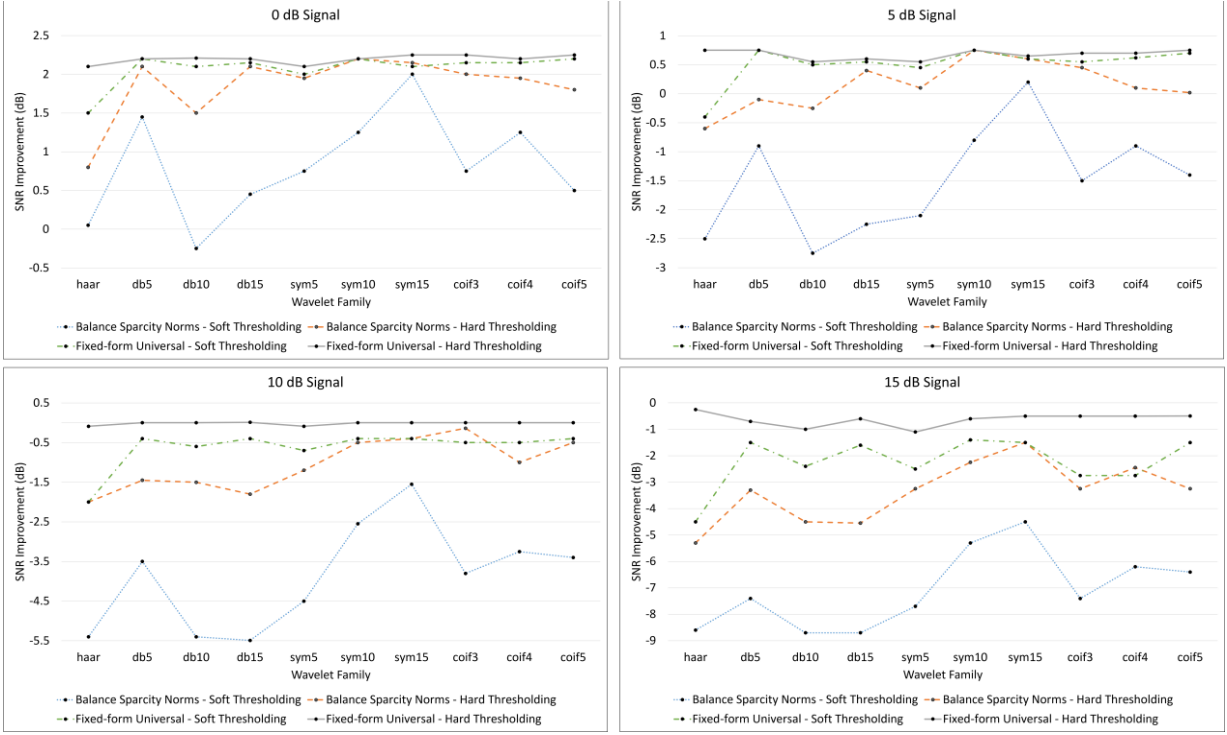

Fig. 4: Output SNR results in wavelet transformation for different SNR values.

$$P(0) = c^{-1}.I \quad (3)$$

$$K(n) = \frac{P(n-1)x(n)}{\gamma + x^T(n)P(n-1)x(n)} \quad (4)$$

$$w_{n+1} = w_n + K(n)e(n) \quad (5)$$

$$P(n) = \frac{1}{\gamma}[P(n-1) - K(n).x^T(n)P(n-1)] \quad (6)$$

where c = 0.99, $\gamma$ = 0.95, P(n) is the correlation matrix, K(n) is the gain vector and order = 80. For AFA,

$$W_{n+1} = \frac{1}{n}\sum_{k=1}^{n} W_k + \frac{1}{n^\gamma}\sum_{k=1}^{n} e(n)x(n) \quad (7)$$

where $\gamma$ = 0.5, order = 450 and for ANLMS,

$$W_{n+1} = \frac{1}{n}\sum_{k=1}^{n} W_k + \frac{1}{n^\gamma}\sum_{k=1}^{n} \frac{\mu}{||x(n)|| \times ||x(n)||}e(n)x(n) \quad (8)$$

where $\gamma$ = 0.05, c = 0.01 and order = 200

*2) Voice Activity Detector (VAD):* We use a VAD algorithm to classify a small segment of an audio signal as voiced or unvoiced. The basic VAD working principle is shown in Figure 3. The input signal is divided into frames to extract features. For the feature extraction, two methods have been used; energy and cepstral analysis. The first few frames are assumed as noise-only frames to get an estimation of the background noise. A threshold value is calculated based on it. Then extracted features from the subsequent frames are compared with the threshold value to obtain the VAD decision. If the features of the input frame exceed the calculated threshold, it declares that speech is present (VAD = 1), and absent (VAD = 0) [16]. As real-world speech signals contain various ackground noises and their signal-to-noise (SNR) is low, it's a disadvantage for an accurate VAD system. In our case, energy-based VAD shows the best results.

*Energy based VAD algorithm*

If the energy of the incoming frame is high, the frame is classified as a voiced frame and an unvoiced frame and vice versa [23]. Here, an adaptive threshold approach was used as described in [16].

*Cepstral based VAD algorithm*

The speech signals $g(n)$ can be expressed as a convolution of sound signal $f(n)$ which contains words and pitch, and the transfer function of the system $h(n)$ which includes the sound quality [12]. Using cepstral analysis, sound signal components can be extracted to give the VAD decision as described in [15].

$$g(n) = f(n) * h(n) \quad (9)$$

C. Datasets

English speech signals from NOIZEUS dataset [21] was used for objective measurement evaluation. It is specifically developed to facilitate the comparison of speech enhancement algorithms among research groups.

For subjective metrics, we recorded a Sinhala language sentence with the consultation of audiologists at the Wickramarachchi Institute of Speech and Hearing. This sentence was chosen to be phonetically balanced and with relatively low word-context predictability. It was recorded in a sound-proof booth, originally sampled at 25 kHz, and down-sampled to 8 kHz. After measuring the active speech level of the signal, scaled bubble noise were added to the speech signal to achieve SNR levels of 0 dB, 5 dB, 10 dB, and 15 dB.

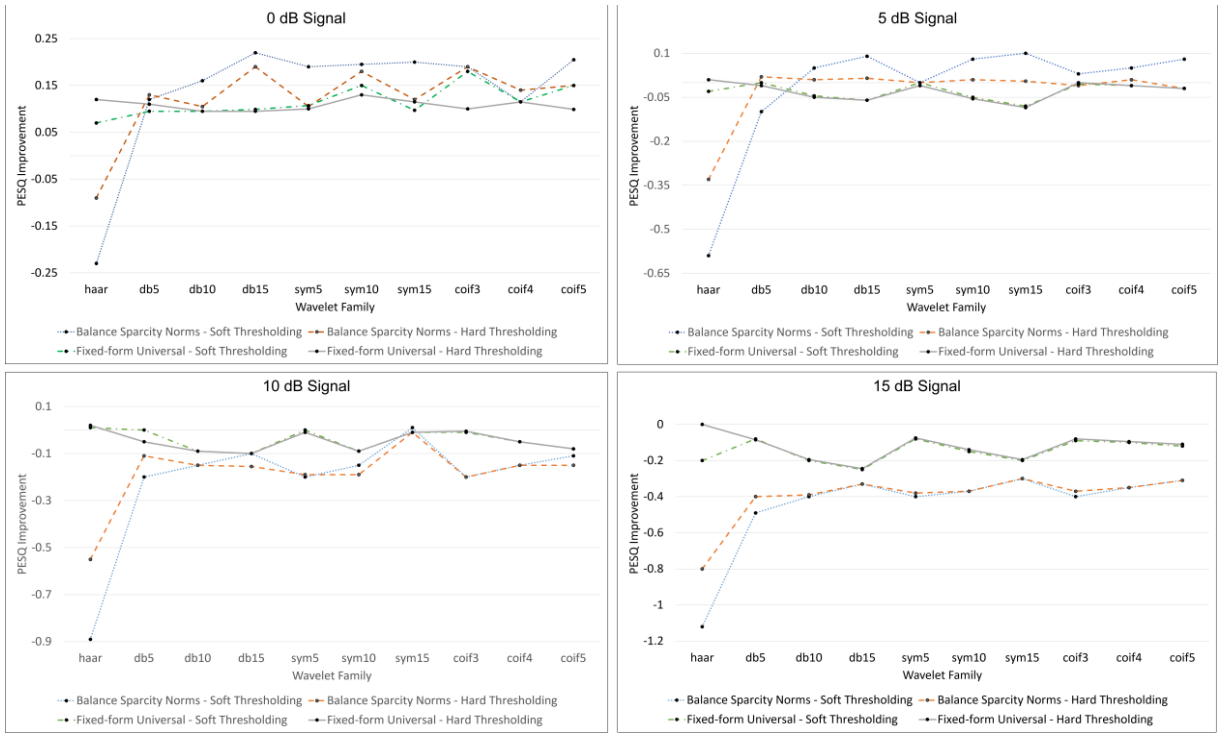

Fig. 5: Output PESQ results in wavelet transformation for different SNR values

### D. Evaluation Metrics

SNR and Perceptual Evaluation of Speech Quality (PESQ) matrices were used as quantitative measurements while Mean Opinion Score (MOS) was used as qualitative measurement. This is due to the contradiction between the results of objective metrics and the user experience on the denoised audio clips. The MOS test was done according to the recommendations in [13], [14], [24].

The MOS is the arithmetic mean over all individual values on a predefined scale that a subject assigns in his opinion of the performance of a system quality. The rating scale maps ratings between worst and Excellent, between 0 and 10 respectively and PESQ is a method to measure the speech quality of an audio signal as recommended in ITU-T P.862 standard. The method has been automated to reflect MOS.

## III. RESULTS

### A. Wavelet Results

After processing through the wavelet algorithms in various combinations, the following results were obtained. When observing output SNR results in Fig. 4, Fixed form universal method with hard thresholding method has yielded the best results.

When SNR value of the input signal is increased, SNR improvement has been decreased. When comparing PESQ results (Figure 5), fixed-form universal threshold with soft method has shown best results in the low SNR values and when SNR value is increased, output PESQ value has been decreased.

### B. Adaptive Filtering Results

Analyzing results from RLS and AFA, LMS, NLMS and ANLMS optimization algorithms suggests that output speech signals are a closer match to the clean signal.

According to Fig. 6, adaptive filters show higher SNR improvements when using the energy-based VAD method. When considering the optimizing algorithms, ANLMS has the highest SNR improvement in both VAD methods. In terms of PESQ results in Fig. 7 it is hard to get a clear idea about the best VAD method. When considering the optimizing algorithms, LMS and NLMS have higher PESQ improvements compared to ANLMS algorithm in both VAD methods.

Using the above results and by listening to the output signals, NLMS and ANLMS adaptive filtering methods along with energy-based VAD were chosen for further testing using the subjective metrics; MOS.

### C. MOS Results

MOS tests were done on 40 normal hearing subjects in the age range of 23 - 27 years. Fig. 8 shows MOS results of six algorithms: NLMS, ANLMS, SYM10, SYM15, NLMS with SYM15, SYM15 with NLMS. According to the results, the adaptive filter using NLMS along with an energy-based VAD shows the best performance for the inputs with high SNR values.

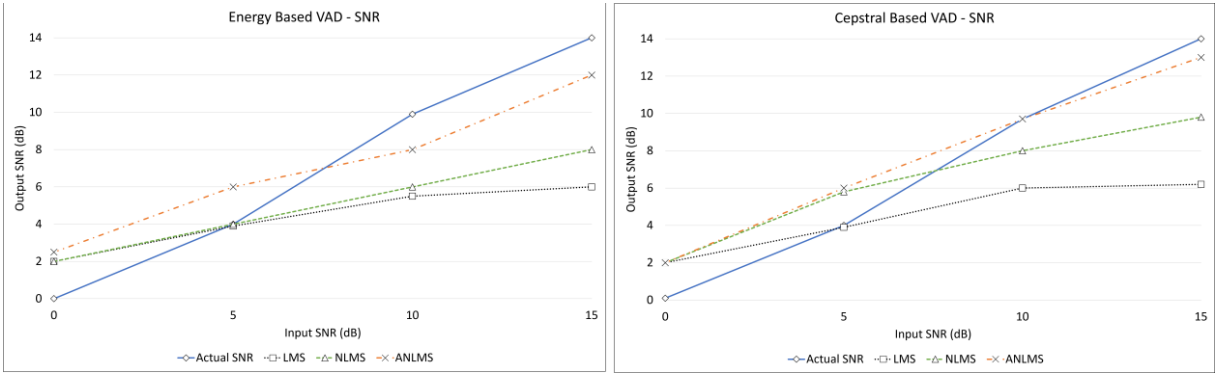

Fig. 6: SNR improvement using energy-based VAD and Cepstral based VAD.

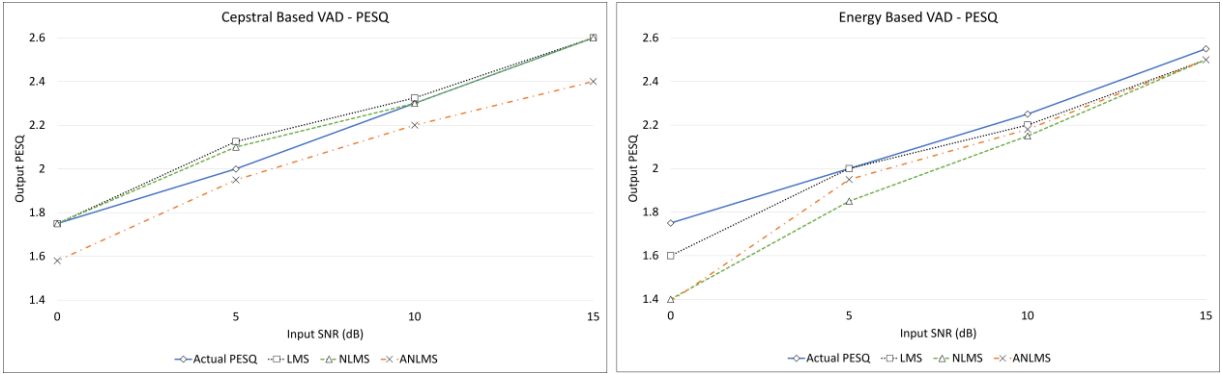

Fig. 7: PESQ improvement using energy-based VAD and Cepstral based VAD

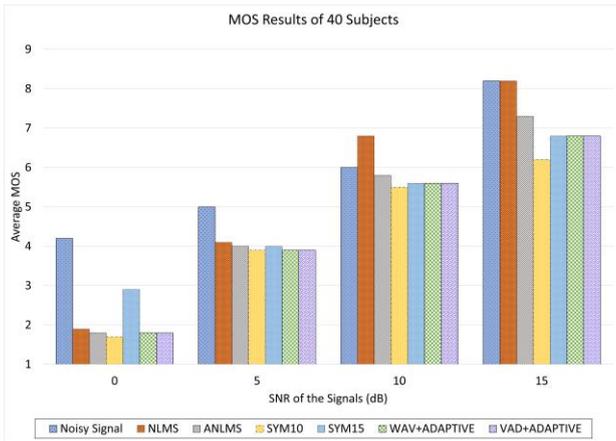

Fig. 8: MOS test results

## IV. DISCUSSION

In this paper, the performance of wavelet transformation and adaptive filtering methods are compared. Two objective metrics (SNR and PESQ) and a subjective metric (MOS) were used for the performance evaluation. The NOIZEUS database is used for the evaluations and a recorded Sinhala language sentence was used for subjective tests.

The use of only one Sinhala sentence may present a limitation due to the rich and diverse structure of the language. Hence, performance results might not be generalized for all Sinhala language processing. Future studies could broaden the scope by utilizing multiple sentences with varied complexity, context, and phonetic structure to achieve more comprehensive results and better understand the behaviour of these algorithms across a broader spectrum.

One notable observation in wavelet transform is that, when SNR value is increased, the performance of the wavelet transform has been decreased. This may be due to the algorithm having suppressed the signal components along with the noise components. Future research could look into this counter-intuitive behaviour and seek to improve the wavelet transform algorithms.

According to the MOS results, and the previous SNR and PESQ results, NLMS adaptive filtering method has shown better performance than the other considered algorithms. While this points towards the potential suitability of NLMS for these types of applications, further research could explore other filtering techniques. Furthermore, an exploration of hybrid techniques, combining the strengths of multiple methods, might yield superior performance.